\documentclass[apj]{emulateapj}
\usepackage{graphicx} 

\shorttitle{Dynamical friction and BSS radial distribution}
\shortauthors{Miocchi et al.}

\def\ltsima{$\; \buildrel < \over \sim \;$}
\def\gtsima{$\; \buildrel > \over \sim \;$}
\def\lsim{\lower.5ex\hbox{\ltsima}}
\def\gsim{\lower.5ex\hbox{\gtsima}}
\def\vv{{\mathbf v}}
\def\rr{{\mathbf r}}
\def\ud{\mathrm{d}}
\begin{document}

\title{Probing the role of dynamical friction in shaping the BSS radial distribution. I
  -  Semi-analytical models and preliminary $N$-body simulations}

\author{
P. Miocchi$^1$,
M. Pasquato$^{2,3}$,
B. Lanzoni$^1$,
F. R. Ferraro$^1$,
E. Dalessandro$^1$,
E. Vesperini$^4$,
E. Alessandrini$^1$,
and Y.-W. Lee$^2$
}

\affil{
\textsuperscript{1} Dipartimento di Fisica e Astronomia, Universit\`a
di Bologna, Viale Berti Pichat 6/2, I-40127, Bologna, Italy\\
\textsuperscript{2} Department of Astronomy \& Center for Galaxy
  Evolution Research, Yonsei University, Seoul 120-749, Republic of
  Korea\\
\textsuperscript{3} Yonsei University Observatory, Seoul 120-749,
Republic of Korea\\
\textsuperscript{4} Department of Astronomy, Indiana University,
Bloomington, Indiana 47405, USA}

\begin{abstract}
We present semi-analytical models and simplified $N$-body simulations
with $10^4$ particles aimed at probing the role of dynamical friction
(DF) in determining the radial distribution of Blue Straggler Stars (BSSs)
in globular clusters. The semi-analytical models show that DF {(which
is the only evolutionary mechanism at work)} is
responsible for the formation of a bimodal distribution with a dip
progressively moving toward the external regions of the
cluster. However, these models fail to reproduce the formation of the
long-lived central peak observed in all dynamically evolved
clusters.
The results of $N$-body simulations confirm the formation of a sharp
central peak, which remains as a stable feature over the time
regardless of the initial concentration of the system.  In spite of a
noisy behavior, a bimodal distribution forms in many cases, with the
size of the dip increasing as a function of time. In the most advanced
stages the distribution becomes monotonic. These results are in
agreement with the observations. Also the shape of the peak and the
location of the minimum (which in most of the cases is within 10 core
radii) turn out to be consistent with observational results.
For a  more detailed and close comparison with observations, including a
proper calibration of the timescales of the dynamical processes
driving the evolution of the BSS spatial distribution,
more realistic simulations will be necessary.
\end{abstract}


\section{Introduction}
\label{Introduction}
Globular clusters (GCs) are dynamically active systems that,
within the time-scale of the age of the Universe,
undergo nearly all the physical processes known in stellar dynamics
\citep{mh97}. Gravitational interactions and collisions among single
stars and/or binaries are quite frequent, especially in the highest
density environments \citep[e.g.][]{hut92}. They can also generate
populations of exotic objects, like X-ray binaries, millisecond
pulsars and blue straggler stars (BSSs; see, e.g.,
\citealp{paresce92,bailyn95,bellazzini95,fe01,ransom05,pooleyhut06,fe09_m30}).

GCs are also old systems where all stars more massive than $\sim 0.8
M_\odot$, the typical main sequence turn-off (MS-TO) mass, should have
already exhausted their core hydrogen reservoir and evolved toward the
sub-giant branch or later phases.  Nevertheless, in all well studied
GCs \citep[e.g.,][]{sandage53,fe92,fe99_m80} BSSs are observed as a
population of core hydrogen-burning stars along an extrapolation of
the MS, in a region of the color-magnitude diagram (CMD) which is
bluer and brighter than the MS-TO.  Their position in the CMD and
direct measurements suggest that these objects are more massive than
the MS-TO stars, with typical masses of $\sim 1.2 M_\odot$
\citep[][]{shara97,gilliland98,demarco05,fe06_COdep,
  lanzoni07_1904,fiorentino13}. To solve this apparent paradox, two
main mechanisms for the formation of BSSs have been proposed, both
involving close physical interactions among stars: mass-transfer in
primordial binary systems \citep{mccrea64, zinn76} and direct
collisions between unbound stars \citep{hills76}.  The two formation
channels could be at work simultaneously within the same cluster,
probably depending on the local density
\citep[e.g.][]{fusipecci92,bailyn92,fe95,fe09_m30}. However, their
relative efficiency is still a matter of debate \citep[e.g.][see also
  \citealt{hypki}]{sollima08,knigge09,chatterjee,sills} and
distinguishing BSSs formed by either of the two mechanisms is a very
hard task.  The only notable exceptions are the detection of
spectroscopic signatures of the mass transfer process in 47 Tucanae
and M30 \citep[][respectively]{fe06_COdep,lovisi13}, and the discovery
of two distinct BSS sequences, likely connected to the two
formation processes,
in M30 and NGC 362 \citep[][respectively]{fe09_m30,ema13_362}.

BSSs are also considered to be powerful probes of GC
internal dynamics
\citep[e.g.][]{bailyn92,fe95,fe99_m80,fe03,fe06_omega,davies04,mapelli04,mapelli06}.
{In particular, \citet{mapelli06} first noted that, in some GCs, the position of
the minimum of the BSS radial distribution nicely corresponds to the radius
where the dynamical friction (hereafter DF) time equals the cluster age.
\citet{fe12} put this observable in an evolutionary context, connecting the
shape of the observed BSS radial distribution with the cluster dynamical age,
thus defining the so-called ``dynamical clock'', a fully empirical tool able
to measure the stage of dynamical evolution reached by these stellar systems.}
In most of the
surveyed GCs, the number of BSSs, normalized to the number of stars in
a reference population (like sub-giants, red giants or horizontal
branch stars), shows a bimodal behavior as a function of radius: it is
peaked in the center, has a dip at intermediate radii, and rises
again in the cluster outskirts \citep[e.g.][and references
  therein]{fe93,lanzoni07_m55,beccari13_5466}.
{A similar behavior has been recently found also in the extra-Galactic
GC Hodge 11 in the Large Magellanic Cloud \citep{li13}.}
  In a few other cases
the BSS radial distribution shows only a central peak
\citep{fe99_m80,lanzoni07_1904,fe09_m30,rodrigo12}, while in
$\omega$ Centauri, NGC 2419 and Palomar 14
\citep[][respectively]{fe06_omega,ema08_2419,beccari11_pal14} it is
equal to the radial distribution of the normal cluster stars.
{Such a flat BSS radial distribution has been found also in dwarf galaxies
\citep{mapelli09,monelli12}.}
Indeed,
the comparative analysis performed by \citet{fe12} in a sample of 21
Galactic GCs demonstrates that these stellar systems can be grouped on
the basis of the shape of their BSS radial distribution, each group
corresponding to families of different dynamical age.  The
interpretative scenario is the following.

Being significantly more massive than normal cluster stars, BSSs are
expected to experience a relatively fast segregation process, mainly
as a ``natural'' consequence of the action of DF, that makes them progressively sink toward the cluster
center.  In general, a ``test'' star of mass $m_\mathrm{t}$, orbiting
at an average radius $r$ in a field of lighter stars with average mass
$\langle m\rangle$ decays toward the cluster center over a time
\begin{equation}
\label{orbitaldecay}
t_{\rm df} (r)\simeq \frac{\langle m \rangle}{m_\mathrm{t}} t_\mathrm{r}(r),
\end{equation}
where $t_\mathrm{r}(r)$ is the relaxation time at the mean orbital radius $r$.
Clearly, once the other parameters are fixed, the larger the value of
$m_\mathrm{t}$, the faster the object sinks to the center. Moreover,
$t_\mathrm{r}$ is expected to increase with
radius, because of its dependence on local density and velocity
dispersion \citep[see, e.g.][]{BT}.  Therefore, heavy stars (as BSSs)
orbiting at large $\langle r \rangle$ will decay extremely slowly,
virtually unaffected by DF (unless they are on very eccentric orbits).
Instead, BSSs that are closer to the center will decay quickly.  On
the other hand, because of their smaller masses, the reference
population stars will be less affected by DF, compared to BSSs.  It is
thus reasonable to expect that, over time, the region in which the
normalized BSS fraction ($n_\mathrm{BSS}/n_\mathrm{ref}$) is depleted
by DF extends increasingly outwards.  In that region, the
behavior of the local BSS fraction exhibits an
absolute minimum (at $r_\mathrm{min}$)
between a central peak (made up of BSSs already decayed, plus
collisional BSSs formed there) and an external rising branch (due to
BSSs that have not had enough time yet to appreciably decay to the
center).

Thus, it is reasonable to expect that in dynamically young clusters
the minimum of the BSS radial distribution should be close to the
center, while for increasing dynamical age, it should be observed at
larger and larger distances.
{Therefore, $r_\mathrm{min}$ can be used as the hand of a ``clock''
able to measure the stage of the dynamical evolution reached by stellar
clusters, with DF being the internal engine of this clock (of course,
for a meaningful comparison among different clusters, $r_\mathrm{min}$
has to be normalized to a characteristic scale length, as the core radius $r_\mathrm{c}$).
Such a tool would also allow to recognize cases where the relaxation process has not
started yet\footnote{Note that
this method, involving relatively bright stars, is much more
effective than any other approach proposed so far to measure the
level of mass segregation (or the lack thereof).}, from those where it
is more advanced, possibly even close to the core-collapse phase.
It may even help to distinguish between systems with a central density cusp
due to core collapse, from those with a cusp due to an intermediate-mass
black hole.}
The empirical indication of the validity of this simplified,
DF-based, picture is provided by the tight correlation found between
the position of the minimum in the observed BSS radial distribution
and the relaxation time computed at the core or at the half-mass
radius \citep[see Figure 4 in][]{fe12}.  The trend has been also
confirmed by additional observational studies
\citep[see][]{ema13_m10,ema13_362,beccari13_5466,sanna14}.

From the theoretical side, Monte-Carlo and $N$-body simulations have
been used to study the radial distribution of BSSs in GCs
\citep{mapelli04,mapelli06,hypki,chatterjee} and binary systems in
open clusters \citep{geller13}. Indeed, these are the main routes to
evaluate the role of DF in shaping the BSS distribution, since they
offer deep insights on the influence of other important collisional
phenomena, like those associated to close encounters and ``binary
burning'' (mainly taking place close to or immediately after the
cluster core-collapse). Therefore, our group is adopting numerical approaches
with gradually increasing levels of realism, in order to precisely
evaluate and disentangle the role of the various dynamical mechanisms
involved. In \citet{alessandrini14} we used a coupled
analytical/$N$-body approach in the specific case of BSSs in a GC
(i.e., test particles only slightly more massive than the average,
orbiting a background field with a mass spectrum), to ascertain that
the observed bimodalities cannot be due to a non-monotonic
radial behavior of the DF time-scale.

Here we first discuss a semi-analytical approach to the problem,
assuming that DF is the only process driving the BSS secular evolution
(Section \ref{semian}). Then, we present collisional $N$-body
simulations to take into account further dynamical mechanisms playing
a role in determining this evolution (Section \ref{Nbody}).  Discussion and conclusions are
presented in Sect. \ref{concl}.

\section{Semi-analytical models}
\label{semian}

\subsection{Basic assumptions}
\label{assumpt}
We neglect BSSs formed through stellar collisions and only deal with
the population generated by mass transfer activity in binary systems.
We further assume that BSS progenitors are dynamically inert hard
binaries, meaning that they suffer only from the effect of DF,
and, moreover, their probability to actually become a BSS is
independent of the cluster environment.  Under these assumptions we
can model the BSS progenitors as point particles with mass equal to
the sum of the binary components. Moreover, we assume that the
progenitors that eventually give rise to BSSs are, at any time, just a
random subsample of the overall progenitor population: hence, at any
time, the radial distribution of these binaries well represents that
of actual BSSs. In other words, it is assumed that BSSs and their
binary progenitors (that we assume as point particles) are
indistinguishable.

Moreover, we consider the cluster as an isolated system, with a
discrete mass-spectrum consisting of only three species meant to
represent MS stars below the TO (the lightest component which
primarily contributes to both the overall gravitational potential
and DF), BSSs
(the most massive component) and the reference population (the
component with intermediate-mass stars and to which BSS star counts
are normalized, corresponding to red-giants or horizontal branch stars
in observational studies).

\subsection{The models}
\label{models}
As a first step in understanding the specific role of DF in shaping
the observed BSS radial distributions, we followed a semi-analytical
approach in which other simplifying assumptions are adopted in
addition to those discussed above.

We considered here the cluster dynamics governed by a static ``mean''
gravitational field ($\mathbf{\Psi}$), as due to MS stars (``field''
stars) only.  We then assumed that $\mathbf{\Psi}$ remains fixed in
time and is generated by a constant, spherically symmetric and
isotropic distribution of field stars, each of which is assumed to
have a mass $m$.  Their phase-space distribution $f(r,v)$ is defined
such that $f(r,v)\ud\rr\ud\vv$ is the number of these stars in the
phase-space volume element $\ud\rr\ud\vv$ (with $r=|\rr|$ and
$v=|\vv|$).

Within this field, we considered the evolution of the stars in the two
heavier components (that we call ``test'' stars) under the effects of
the DF against field stars. We neglected the self-gravity acting on all
these components, as well as any interaction between test
stars. Hence, the stellar motion of any test star, with mass
$m_\mathrm{t}$, position $\rr$ and velocity $\vv$, is determined
solely by the underlying gravitational field $\mathbf{\Psi}$ and by
the DF deceleration that we describe following the \citet{chandra}
formula \citep[see, e.g.,][]{BT}
\begin{equation} 
{\mathbf a}_\mathrm{df}=-4\pi\ln\Lambda G^2m(m+m_\mathrm{t})g(r,v)v^{-3}\vv,
\label{DF} 
\end{equation}
where
\begin{equation}
g(r,v)\equiv 4\pi\int_0^v f(r,w) w^2\ud w
\label{grv1} 
\end{equation}
is the number density at radius $r$ of field stars moving slower than
the considered test star. Assuming that field stars are distributed 
according to {the} \citet{plummer} distribution function with a
{fixed} scale length $r_0$, an
analytical expression (see Eq. [\ref{grv}] in the Appendix) can be
derived for $g(r,v)$, while the gravitational field is given by
\begin{equation} 
\mathbf{\Psi}(\rr)=-\frac{G M}{(r^2+r_0^2)^{3/2}}\rr,
\label{afield}
\end{equation}
with $M$ being the total mass in the field star component.  Thus, once
numerical values for $m$, $m_\mathrm{t}$, $M$ and $r_0$ are chosen,
the DF deceleration acting on BSSs and the reference
stars is completely determined.  In particular, we assigned to BSSs a
mass $3m$ and to the reference population stars a mass $2m$.  In
physical units, this choice can be thought to correspond to $m=0.4~
M_\odot$ for the mean stellar mass below the MS-TO, $0.8~M_\odot$ for
stars in the reference population and $1.2~M_\odot$ for BSSs, all
being appropriate values for the case of Galactic GCs.

As initial conditions for the time evolution of the two evolving
components, we generated a set of $N_{\rm BSS}=300$ and a set of
$N_{\rm ref}= 1200$ positions and velocities for their representative
particles, following the same Plummer distribution function used for
the field stars.
{Of course, in real clusters the relative abundance of BSSs with
respect to the reference population is much lower than it is assumed here.
However such a large number of BSSs is adopted to limit the Poisson noise.
To this end, we also generated $20$ sets of initial conditions by
changing only the random seed, and we then merged the snapshots of
the resulting simulations, after having reported the center of mass
of each snapshot onto the origin of the coordinates. In addition, for
each particle in each snapshot, we merged the three projections
(along each coordinate axis), thus obtaining (from a statistical point
of view) three times more stars. In this simplified approach, such an
overabundant BSS population has no consequences on the system evolution,
while it gives some spurious effects in the $N$-body simulations, as
we will discuss later in Sect.~\ref{setup}.}
The assumption of the same initial distribution function for both
kinds of particles and the field component is empirically justified by
observations: in fact, BSSs are found to share the same radial
distribution as normal cluster stars in dynamically young GCs, where
DF has not been effective yet in segregating massive stars toward the
cluster center \citep[see the cases of $\omega$ Centauri, NGC 2419 and
  Palomar 14 in][and references therein]{fe12}.

Starting from these initial conditions, the orbit of each test star,
evolving under the total acceleration ${\mathbf
  a}=\mathbf{\Psi}+{\mathbf a}_\mathrm{df}$, was time-integrated by
means of a 2nd order leapfrog algorithm \citep[e.g.][]{HE} with
constant time step. At given times, a snapshot of the system was
extracted and the projected number distribution of the two heavier
stellar species was derived in a series of concentric annuli around the
cluster center. To further improve the statistics we superimposed the
positions of the test particles in all the 20 realizations, as well as
their projections on the three coordinate planes (similarly to what
done in the $N$-body model, see Section \ref{Nbody}).

\subsection{Results}
\label{resu_semian}
\begin{figure}
\includegraphics[width = \columnwidth]{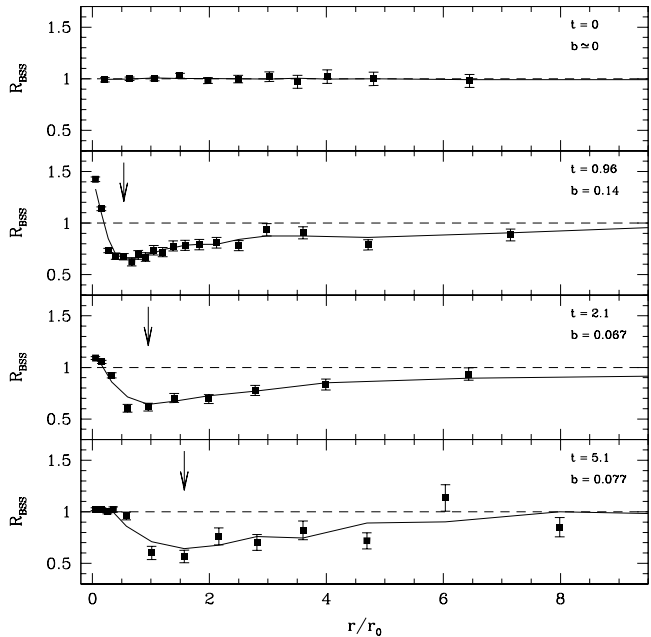}
\caption{Double-normalized ratio ($R_\mathrm{BSS}$) between the
  projected number of BSSs and that of reference stars, in various
  radial bins around the cluster center and at different evolutionary
  times (see labels), as found from the semi-analytical
  simulations. Time is expressed in units of the half-mass relaxation
  time $t_\mathrm{rh}$. The solid curve is the running average
  of the $R_\mathrm{BSS}$ radial behavior.  Also labeled is the slope
  ($b$) of the rising branch beyond the dip (see text). The number of
  radial bins is variable due to the employed adaptive binning method.
}
\label{nratio}
\end{figure}
Consistently with the observational quantities defined in
\citet{fe93}, in Figure \ref{nratio} we plot the ``double-normalized''
BSS radial distribution
$R_\mathrm{BSS}(r)\equiv(n_\mathrm{BSS}(r)/N_\mathrm{BSS})/(n_\mathrm{ref}(r)/N_\mathrm{ref})$,
i.e. the ratio between the relative fraction of BSSs and that of
reference stars in each radial bin, at various evolutionary times.
The radial distance is expressed in units of $r_0$, while times are in
units of $t_\mathrm{rh}$, namely the relaxation time computed at the
half mass radius \citep[][Eq. 8-72]{BT} of the field star system:
$r_h=r_0(2^{2/3}-1)^{-1/2}\simeq 1.3 r_0$ for the Plummer
distribution.

To guarantee both a good radial sampling and a large enough number
statistics, we set a minimum threshold ($n$) to the number of BSSs and
a minimum threshold ($\Delta r$) to the width of each radial
bin. Then, the actual width of each bin was automatically determined as
the minimum value larger than $\Delta r$ such that $n_\mathrm{BSS}\geq
n$ in that bin.  In Fig.~\ref{nratio} we chose $n=300$ and $\Delta
r=0.1 r_0$ for all times, except at late stages ($t= 5.1$) where
$n=100$ was used in order to get enough resolution also in the outer
regions where the number of BSSs is quite small, and for $t=0$ where
$\Delta r=0.4 r_0$.  Uncertainties on the number ratios were
estimated from the law of propagation of errors, assuming a Poissonian
statistics for $n_{\rm BSS}$ and $n_{\rm ref}$.

\begin{figure}
\includegraphics[width = \columnwidth]{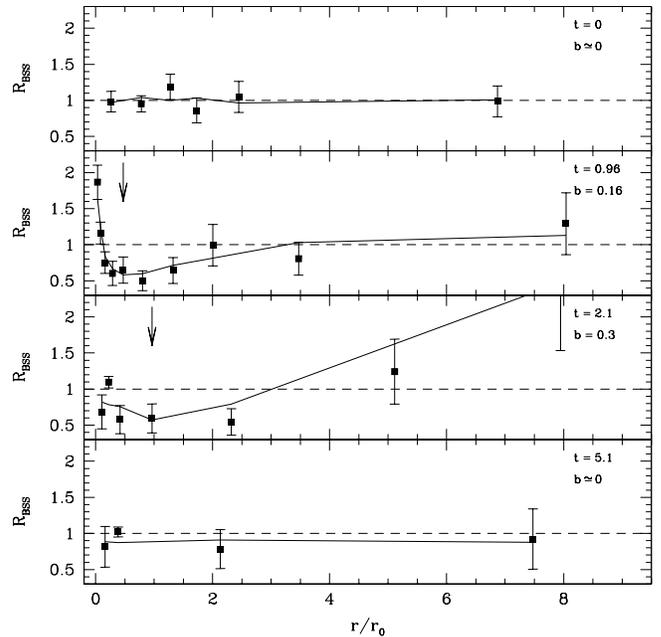}
\caption{{The same as in Fig.~\ref{nratio}, but considering
the snapshots of only one simulation and with no merging of coordinate planes.}
}
\label{nratio_noav}
\end{figure}
{For illustrative purposes, Figure \ref{nratio_noav} shows the
$R_\mathrm{BSS}(r)$ profiles obtained by considering only one simulation in one projection direction.
For obvious reasons, in doing this plot
we had to reduce the BSS number bin threshold, $n$, by a factor $\simeq 10$
and $\Delta r$ by a factor of a few times.
Unsurprisingly, with respect to Fig.~\ref{nratio}, it can be seen a strong increase
of the Poisson noise,
especially in the outer regions where the number of particles is the lowest. 
This example illustrates the importance of reducing counting noises,
as done in applying the averaging procedure described above.}

In order to properly characterize the shape of the $R_\mathrm{BSS}$
distribution and the location of its minimum ($r_\mathrm{min}$) in
each snapshot, we first computed the running average\footnote{It is a simple and
central moving average with a window width of 3 bins.} $\langle
R_\mathrm{BSS}(r)\rangle$, with the aim of reducing fluctuations due
to poor statistics. Then, $r_\mathrm{min}$ was defined as the distance
from the cluster center of the absolute minimum of this average.
While a flat behavior is set by construction at the initial time, a
bimodality rapidly develops {(see Fig.~\ref{nratio})}.
Moreover, the minimum of the normalized
BSS radial distribution progressively drifts outward at increasing
evolutionary times {(note that $r_0$ is constant by construction
in these Plummer models)}, until an almost flat behavior is established at
late stages ($t\gtrsim 10 t_\mathrm{rh}$).  These results
qualitatively confirm what suggested by the intuitive picture
discussed above, namely that DF by itself can give rise to a bimodal
BSS distribution. {On the other hand, we note that although
a central peak begins to develop from the very beginning,
it is then rapidly leveled-off, at odds with what
observed in real clusters. This is because, in this simplified model,
the frictional decay of both kinds of test stars continue
indefinitely with an unaltered efficiency, thus making
the great majority of these stars to eventually decay to the
innermost radial bins where, as a consequence, the peak in
$n_{\rm  BSS}/n_{\rm ref}$ is dumped to nearly its initial value
(i.e.~$R_{\rm BSS}=1$).
Indeed, the innermost 5 bins ($r/r_0<1$) in the bottom panel of Fig.~\ref{nratio}
contains $\sim 95$\% of the total test stars.}

To quantify the level of bimodality of the distribution, we defined the
parameter $b$ as the slope of the line that best-fits $\langle
R_\mathrm{BSS}(r)\rangle$ in the region of the rising branch
(specifically for $r_\mathrm{min} \leq r \leq r_\mathrm{min} + 4\Delta
r$). A visual inspection of the snapshots indicates that the
dip in $R_\mathrm{BSS}(r)$ can be well appreciated when $b\gtrsim
0.01$.\footnote{We note, however, that an automatic parametrization of
the bimodality is not an easy task, since the shape of the region
where the minimum of the distribution is located significantly
changes with time. In particular, at late stages of the cluster
evolution this region broadens and the $b$ parameter tends to become
less sensible and easily lose the bimodal behavior.}

\begin{figure}
\includegraphics[width =\columnwidth]{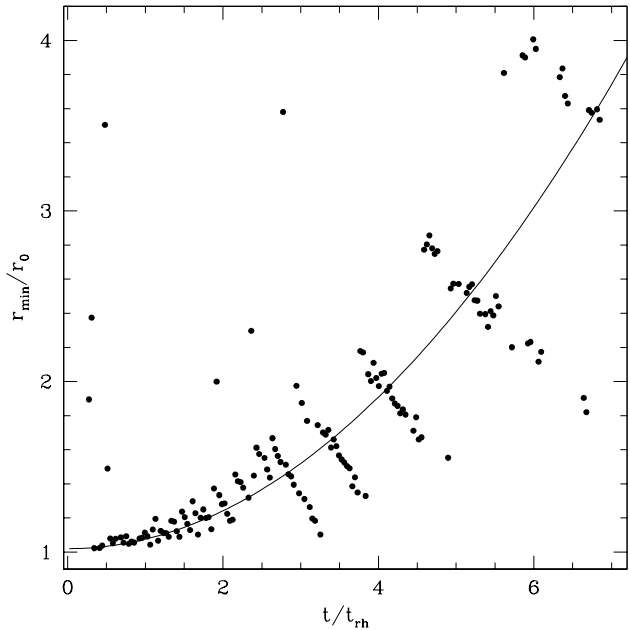}
\caption{Time evolution of the position of the absolute minimum,
  $r_\mathrm{min}$ (dots), normalized to the Plummer characteristic scale
  radius, for all the snapshots in which $R_\mathrm{BSS}(r)$ has an
  appreciable bimodality ($b\gsim 0.01$; see text).  The solid line is
  a fitting $\sim t^2$ law.
}
\label{rmin_semian}
\end{figure}
The time behavior of $r_\mathrm{min}$, as measured in all
distributions having $b\geq 0.01$, is plotted in Figure
\ref{rmin_semian}.  From the inspection of this Figure, we can state
that (despite the large fluctuations): $(i)$ the $R_\mathrm{BSS}$
radial behavior shows a significant level of bimodality most of the time
for $ 0.5 \lesssim t/t_\mathrm{rh} \lesssim 7$; $(ii)$ there is a
clear tendency of $r_\mathrm{min}$ to drift outward.
The gaps among groups of $r_\mathrm{min}$ values
shown in this Figure (e.g.~between $t/t_\mathrm{rh}\simeq 4.2$ and $\simeq 5.5$),
as well as the linear anti-correlation within these groups, are an effect
of the adaptive binning procedure.

\section{$N$-body simulations}
\label{Nbody}
To get deeper insights into the role played by DF and, possibly, other
collisional effects on the observed shape of the BSS normalized radial
distribution, we followed a more realistic approach making use of
self-consistent, collisional $N$-body simulations. The same basic
assumptions outlined in Sect. \ref{assumpt} have been adopted\footnote
{While only \emph{single} stars are generated in the
initial conditions of the simulations, binary and multiple systems
can form dynamically during the evolution.}.  Nevertheless, here we have
an accurate and self-consistent dynamical treatment of the various
fully interacting stellar components, naturally including DF and close
encounters, which are responsible for various dynamical phenomena
\citep[e.g.][]{mh97,mbp}.

The simulations were performed using the direct $N$-body code
\texttt{NBODY6} \citep{aar03} with its Graphic Processing Unit
extension enabled.  We adopted the ``H\'enon units'' (also known as $N$-body units)
discussed in
\citet{hm86}, where $G=M=-4E=1$, with $G$ being the gravitational
constant, $M$ the total GC mass and $E$ the total GC energy (the sum
of potential and kinetic energy, negative for a bound system). In
these units, the half-mass relaxation time is \citep{giersz94}:
\begin{equation}
\label{trh}
t_\mathrm{rh} = \frac{0.138 N r_h^{3/2}}{\ln{(0.11 N)}} 	
\end{equation}
where $N$ is the total number of particles (stars).  While
$t_\mathrm{rh}$ varies during the evolution because of the changes in
both $r_h$ and $N$ \citep[a star is removed from the system when its
  total energy is positive and it is outside $10 r_h$;
  see][Sect.~9.6]{aar03}, in the following the time will be measured in
units of $t_\mathrm{rh}$ as evaluated from the initial conditions (at
$t=0$).  Note, finally, that, due to the freedom of scaling the
simulation from H\'enon to physical units, only the mass- and the
number-ratios of the species are relevant to the dynamics of the
system, given the total number of stars.

\subsection{Setting up the simulations}
\label{setup}
We fixed the total number of stars to $N=10^4$. Moreover, the same
initial conditions were adopted for the three mass components: at
$t=0$ they all follow a \citet{king66} model with the same central
dimensionless potential $W_0$ and King radius.  As in the
semi-analytical models, this corresponds to assuming no initial mass
segregation and an observationally-justified flat radial distribution
for the ratio between the number of BSSs and that of reference stars
initially. To check for possible dependences of the results on the
initial cluster concentration, we ran three sets of simulations for
three different values of $W_0$, namely $W_0=4, 6, 8$ (corresponding
to King concentration parameters $c\sim 0.84, 1.2, 1.8$,
respectively).

As in the semi-analytical model, we assumed the reference population
stars to have a mass $m_{\rm ref}=2m$, and BSSs having a mass $m_{\rm
  BSS}=3m$. The relative number of the three species is more tricky,
since real cluster's BSSs are numerically negligible with respect to
the other two populations, but here high-enough statistics is needed to
obtain meaningful results. We therefore assumed $N_{\rm MS}=8500$ and
(as in Section \ref{semian}) $N_{\rm ref}=1200$ and $N_{\rm  BSS}=300$,
so as to get a reasonable compromise between realistic values and a
good statistical sampling. {As done in the model of Sect.~\ref{semian}, also
in this case we generated $20$ sets of initial conditions
and merged the snapshots of
the resulting simulations, as well as the projections on the
coordinate planes, so as to obtain, at any sampled
time,} a ``super-snapshot'' made up of $20\times 3\times 10^4=6\times 10^5$
particles\footnote{This number actually refers to the beginning of the
  simulations, for a certain fraction of stars escapes from the system
  during its evolution (because of evaporation and/or ejection).}.
{The unrealistically high fraction of BSSs increases the collisionality
of the system in the central region, with the effect of
making the evolution toward the core-collapse faster.
Nonetheless, in this preliminary study
we preferred to keep the statistical fluctuations low,
even at the price of assuming a less realistic fraction of BSSs. 
However, in order to determine the importance of the enhanced collisional effect
that this choice implies, an} additional,
more realistic, cluster model with $N=10^5$ (and $N_{\rm
  ref}=3000$, $N_{\rm BSS}=300$) was also considered but, due to the
huge computational costs, no statistical sampling was possible for the
initial conditions and only one simulation (with $W_0=8$)
was performed in this case.

All simulations were run for several initial half-mass relaxation
times or until \texttt{NBODY6} failed to meet the desired minimal
energy conservation accuracy \citep[we set the $Q_\mathrm{E}$
  parameter in \texttt{NBODY6} input file to $10^{-4}$; see][]{aar03}.
We then extracted a snapshot per H\'enon unit of time. Since the
typical crossing time of our models is of the order of several
H\'enon time units, this ensures that the evolution of individual
star orbits is tracked in a relatively fine-grained way.

\subsection{Results}
\label{remarks}
\begin{figure}
\includegraphics[width =\columnwidth]{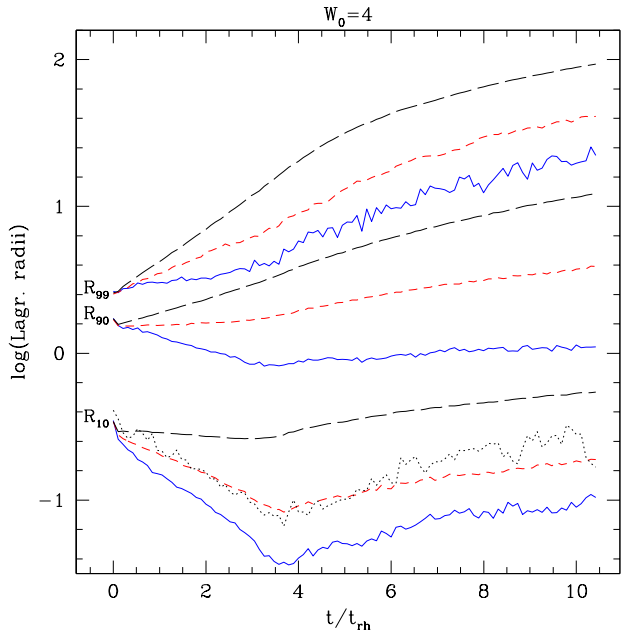}
\caption{Evolution of three Lagrangian radii, $R_p$ (expressed in
  H\'enon units), enclosing the indicated percentage, $p$, of the
  total mass, for the three stellar components in the simulations
  started with $W_0=4$: BSSs (solid line, colored blue in the online version); reference component
  (short-dashed line, red in the online version); MS stars (long-dashed line). The
  dotted line represents the behavior of the core radius of the
  reference component, $r_\mathrm{c}(t)$.  For the sake of clarity, one every
  20 snapshots ($\sim t_\mathrm{rh}/10$) are considered in this plot.
}
\label{rlags1_4}
\end{figure}
\begin{figure}
\includegraphics[width =\columnwidth]{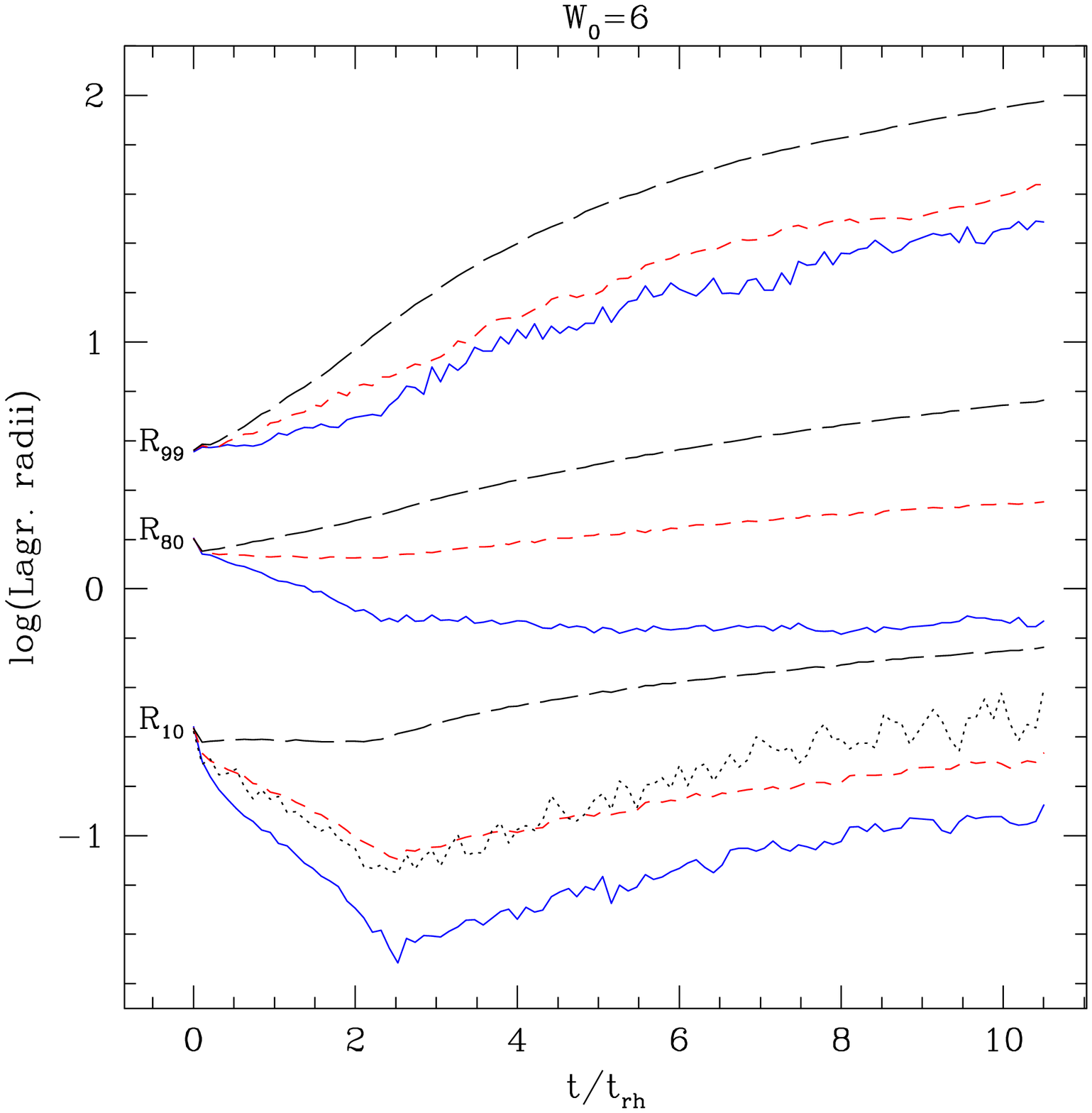}
\caption{As in Fig. \ref{rlags1_4} for the simulations started with
$W_0=6$.}
\label{rlags1_6}
\end{figure}
\begin{figure}
\includegraphics[width =\columnwidth]{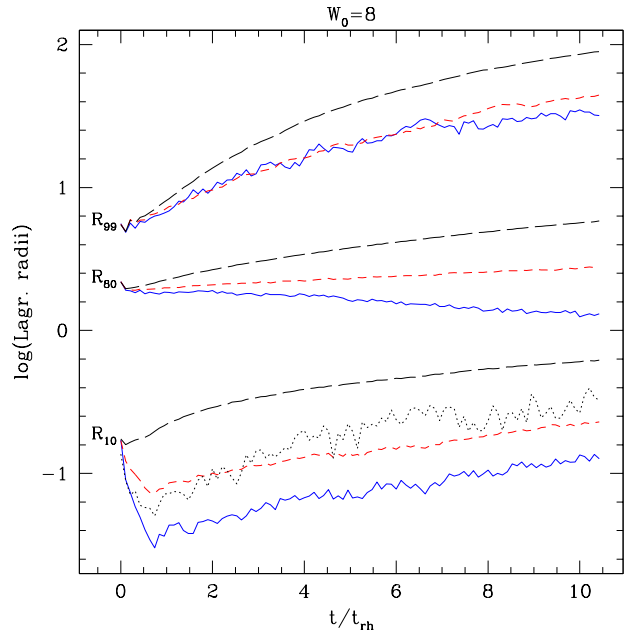}
\caption{As in Fig. \ref{rlags1_4} for the simulations started with
  $W_0=8$.}
\label{rlags1_8}
\end{figure}
\begin{figure}
\includegraphics[width =\columnwidth]{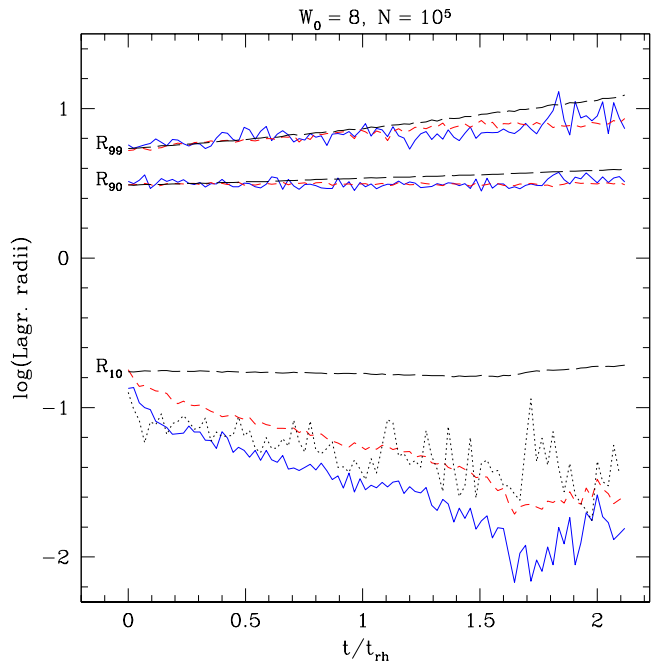}
\caption{As in Fig.~\ref{rlags1_8} for the simulation with $N=10^5$
  stars, of which $N_\mathrm{BSS}=300$, $N_\mathrm{ref}=3000$.}
\label{rlags_W8_100k}
\end{figure}

Before going further in describing the BSS radial distribution
resulting from the $N$-body approach, it is worth analyzing the
overall evolution of the simulated stellar systems.  In Figures
\ref{rlags1_4}--\ref{rlags1_8} the evolution of three representative
Lagrangian radii ($R_p$, with $p$ being the percentage of the total
mass they enclose in 3-D) is reported for the three mass components
and for each of the considered $W_0$ values, while Fig.~\ref{rlags_W8_100k}
refers to the $N=10^5, W_0=8$ case.  We chose: $R_{10}$, that
roughly measures the size of the core region, $R_{99}$, that
corresponds to the very outer halo, and an intermediate Lagrangian
radius delimiting a region of the cluster not participating to the
late expansion phase of the system.  Their behaviors
essentially confirm what expected from the known collisional
relaxation processes in multi-mass systems, of which extensive
descriptions can be found in the literature \citep[e.g.][and
  references therein]{mbp, gurkan04, khalisi}. Here, it is worth
pointing out that the heavier components evolve toward a
core-contraction ($R_{10}$ decreases), while the halo ($R_{99}$)
monotonically expands, with the evolutionary time scale being shorter
for the heavier mass components (cf. eq. [\ref{orbitaldecay}]).  The
halo expansion occurs mainly because the kinetic energy of
(dynamically ``cold'') halo stars increases during \emph{close}
encounters with (``hot'') core stars, especially in the central denser
region. Incidentally, this mechanism also explains the apparent lack
of core-contraction for the (lighter) field stars (their $R_{10}$ is
never contracting), which is due to the injection of kinetic energy
from the contracting cores of the heavier components.

In the same figures we also report (as dotted lines) the time evolution
of the core radius of the reference population.  To be as close as
possible to the observational procedures adopted for real clusters, we
searched for the best-fit King model to the central portion of the
projected number density profile of the reference population and we
defined $r_\mathrm{c}$ as the radius at which the surface density drops to half
its central value.  This well corresponds to the core radius adopted
in observational works and it allows a meaningful comparison among BSS
radial distributions determined in different GCs \citep[see][]{fe12}.
The behavior of $r_\mathrm{c}$ is close to that of $R_{10}$: it shows a
well defined contraction phase, followed by an expansion.
The relatively sudden turnaround of $r_\mathrm{c}$ and $R_{10}$ flags the onset of
the so-called the core-collapse (CC) process.

It can be seen that in our
$10^4$ particles $N$-body simulations the CC phase starts at
$t_\mathrm{CC}/t_\mathrm{rh}\simeq 3.7, 2.5, 0.7$ for $W_0=4,6,8$,
respectively. As expected, the CC time is anti-correlated with the
initial cluster concentration (i.e., with the collision rate in the
core; see the ``heavy tracers'' case in \citealp{fregeau02}).
However, we point out that the particular values of $t_\mathrm{CC}/t_\mathrm{rh}$
are not meant to be used for a close comparison with observational data
since they are specific to the simplified initial conditions adopted here. In fact,
the evolution of the simulated systems is unrealistically
influenced by the heaviest components which are largely overabundant
($N_{\rm BSS}/N_{\rm ref}\sim 0.25$ and $N_{\rm ref}/N_{\rm MS}\sim
0.14$) with respect to reality. Indeed, it is well known that, in
general, the higher these ratios, the faster the collisional
relaxation and the earlier the CC time, compared to the
single-component case \citep[see, e.g., the comprehensive discussion
  in Sect.~1.2 of] [and references therein; see also Table 2 of
  \citealt{fregeau02}]{gurkan04}.  Thus, it is reasonable to expect
that in a real GC, where the total stellar mass in BSSs and reference
populations relative to the total cluster mass is much lower than in
our $N$-body models, the evolution is comparably closer to that of a
single-component system, characterized by a later core-contraction
phase.

This is indeed confirmed by the $N=10^5$ simulation results,
where the number ratios among the different populations are more
realistic. As shown in Figure \ref{rlags_W8_100k}, the CC time in
this simulation is increased by a factor 2.3
($t_\mathrm{CC}/t_\mathrm{rh}\simeq 1.6$ for $W_0=8$) with respect to
the $10^4$ particle case. This comparison clearly shows that the
$10^4$ particle simulations presented here are too rough to provide
accurate estimates of the characteristic time-scales of the various
dynamical processes. However, they can be used to investigate
interesting trends and draw qualitative conclusions.  The analysis of
the trends shown in Figs. \ref{rlags1_4}--\ref{rlags1_8} is indeed
quite instructive.

In particular, the behavior of the Lagrangian radii as a function of
time nicely highlights the properties of environmental conditions in
which DF operates in real clusters. In fact, at odds with the static
environment considered in the semi-analytical model, real clusters have
time evolving environments where DF drifts heavy stars toward the center,
first, in a contracting core (until the CC occurs), and then in an
expanding core (after the CC).  Thus, in the late evolutionary stages,
DF can be somehow contrasted by the core expansion. However, it is
worth noticing that $R_{10}$ for the BSS population is significantly
smaller that the typical size of the central peak in observational
studies.  Hence its time behavior (which is qualitatively similar to
that of $R_{10}$ and $r_\mathrm{c}$ for the reference population) is not
expected to have a significant impact on the overall shape of the BSS
distribution, apart from a possible increase of the width of the peak
and a stabilization of its height in the post-CC regime.

\subsection{Formation of the bimodal behavior}
\label{resu}
Within the ``evolutionary'' picture described above, we now examine the
$R_\mathrm{BSS}$ profiles of the simulated $N$-body systems and
compare them to what obtained from the semi-analytical model and the
observations.
The single super-snapshot shows a noisy behavior,
hindering the automatic analysis of the BSS distribution,
which risks to lose important features (such as the depth of the
minimum) and to fail a reliable evaluation of the bimodality.
For this reason we used the
adaptive binning procedure described in Sect.~\ref{resu_semian} to build
the $R_\mathrm{BSS}$ profiles. {The prescriptions adopted to determine
the location of the minimum ($r_\mathrm{min}$) and to evaluate the bimodality level ($b$) can also
be found in that Section.}
According to the observations, the distance from the cluster center was scaled to
the instantaneous value of $r_\mathrm{c}$ computed as described above.

\begin{figure}
\includegraphics[width =\columnwidth]{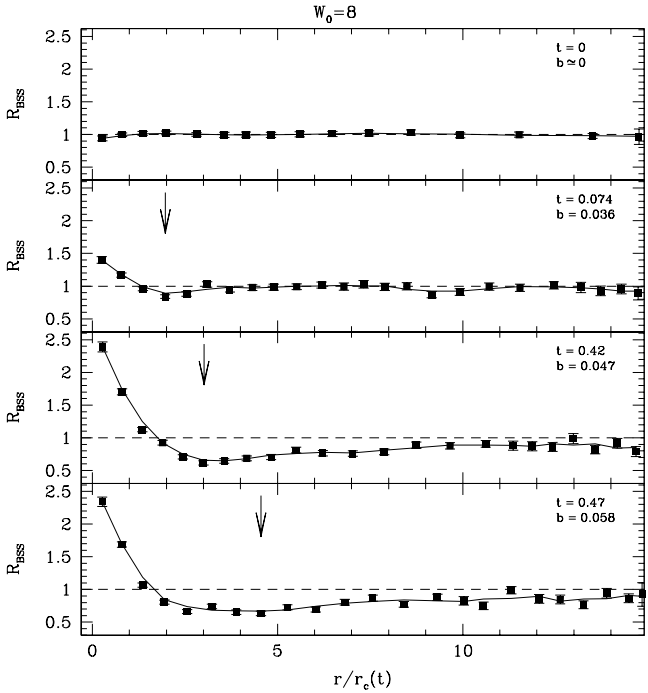}
\caption{$R_\mathrm{BSS}$ profile (dots) for the $N$-body simulations
  starting with $W_0=8$, at different evolutionary times (see labels).
  Dashed line: initial (unsegregated) value of $R_\mathrm{BSS}$.  The
  solid curve is the running average of $R_\mathrm{BSS}$ and the
  bimodality indicator ($b$) is also reported.  The arrow marks the
  location of the absolute minimum, $r_\mathrm{min}$.}
\label{ratio_8}
\end{figure}
\begin{figure}
\includegraphics[width =\columnwidth]{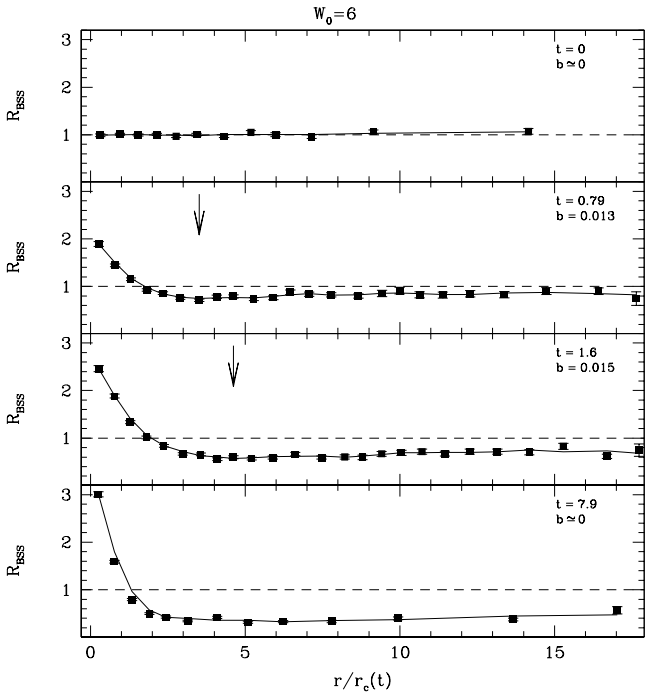}
\caption{As in Fig. \ref{ratio_8} for the $N$-body simulations
  starting with $W_0=6$.}
\label{ratio_6}
\end{figure}
\begin{figure}
\includegraphics[width =\columnwidth]{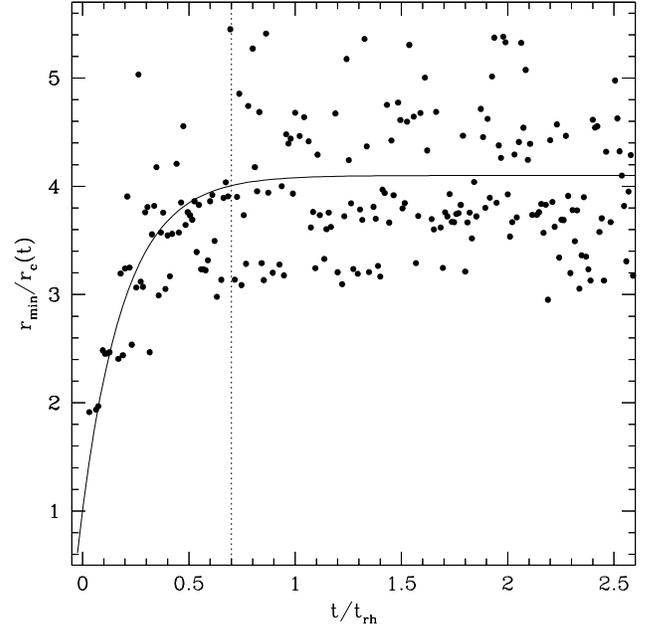}
\caption{Time evolution of the position of the absolute minimum of the
  $R_\mathrm{BSS}$ profile (dots), in units of the instantaneous core
  radius, for the simulations started with $W_0=8$. A fitting exponential law ($\sim
  1-\exp(5t/t_\mathrm{rh})$) is shown as a solid curve.  Only
  $R_\mathrm{BSS}$ profiles for which $b>0.01$ have been considered
  (see text).  The dotted line indicates $t_\mathrm{cc}$, the time of
  the core-collapse.}
\label{rmin_t}
\end{figure}

In the lowest concentration case ($W_0=4$, not shown here) a central peak in the BSS
radial distribution is soon developed and the external portion of
$R_{\rm BSS}(r)$ rapidly decreases without forming a significantly
bimodal pattern.
{This can be understood by the fact that the time-scale
of the frictional decay depends predominantly on the density of the field stars
\citep[e.g.][]{alessandrini14}, and this density
decreases more gradually with radius in low concentration clusters than in
those highly concentrated.
Hence, in the case $W_0=4$ the DF time-scale increases more slowly outward
(i.e., its efficiency keeps relatively high up to a larger radius)
than for $W_0=6, 8$.
This can be seen in Fig.s \ref{rlags1_4},
\ref{rlags1_6} and \ref{rlags1_8}, by comparing the slope of
the inner Lagrangian radius behavior with that of the intermediate
radius for the BSSs in the pre-CC phase.
It is evident that in the $W_0=4$ case
the trends of this two radii show more similar slopes than
for $W_0=6, 8$.
This means that in the $W_0=4$ model the BSSs decay at a rate
that is almost independent of the radius (at least up to $R_{90}$), thus  
making the double-normalized ratio to evolve very quickly towards
the unimodal pattern. In fact, in order to ensure the persistence of
a bimodal distribution, the BSSs orbiting in the outskirts have to
decay much more slowly than those orbiting in the inner region.}

Representative examples of the radial distributions
obtained in simulations with initial potential $W_0=8$ and $W_0=6$ are
shown in Figures \ref{ratio_8} and \ref{ratio_6}, respectively, for
the labeled evolutionary times (in units of the initial
$t_\mathrm{rh}$).  By construction, $R_{\rm BSS}(r)$ is nearly
constant and close to the unity at the initial time.  As the evolution
proceeds, a bimodal behavior develops, with an increasingly high
central peak and a dip at intermediate radii (see also Fig.~\ref{rmin_t}).
A number of interesting features can be inferred from these simulations:
\begin{enumerate}
\item all snapshots show the formation of a sharp central peak in the
  BSS radial distribution, regardless of the initial value of $W_0$
  (including $W_0=4$);
\item at odds with the findings of the semi-analytical models the central 
peak is a quite stable feature;
\item the number of BSSs drifted to the center, because of the effect
  of DF, increases as a function of time;
\item in many cases the BSS distribution is bimodal (Fig.~\ref{rmin_t}). This effect is
  somehow mitigated by a progressive decrease of $N_{\rm BSS}$ in the
  external regions, which makes the detection of bimodality difficult
  and needs to be further investigated;
\item the width of the dip seems to increase as a function of time, in
  nice agreement with the observations;
\item in the latest snapshots, the simulated BSS distribution shows a
  monotonic behavior, with most of the BSSs segregated in the central
  part and the external radial bins being essentially devoid of BSSs
  (see the bottom panel of Fig.  \ref{ratio_6}), in agreement with the
  BSS distributions observed in Family III clusters \citep{fe12};
\item in the cases where the bimodality is clearly distinguishable, an
  outward drift of the dip for increasing evolutionary time is seen
  before the CC phase (see Figure \ref{rmin_t}).
\end{enumerate}
Remarkably, the shape and width of the central peaks in Figs.
\ref{ratio_8} and \ref{ratio_6} are also very similar to those
observed in real clusters belonging to Family II (i.e., those actually
showing a bimodal BSS distribution; see \citealp{fe12}). In fact, as
apparent in their Figure 2, the large majority ($\sim 85$\%) of these
systems have $r_\mathrm{min}$ smaller than $10 r_\mathrm{c}$, consistently with
the results of both our simulations with concentrated initial
conditions and with the semi-analytical model results.

\begin{figure}
\includegraphics[width =\columnwidth]{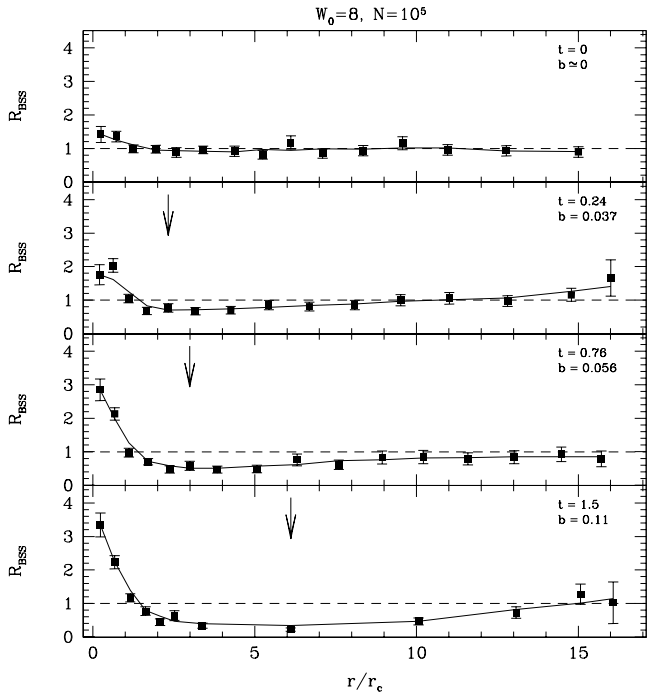}
\caption{{$R_\mathrm{BSS}$ profiles at different evolutionary times
for the simulation run with $N=10^5$ and $W_0=8$.
Symbols and labels are as in Fig.~\ref{ratio_8}.}}
\label{ratio_8_100k}
\end{figure}
{These results seem to be confirmed (at least qualitatively)
by the $R_\mathrm{BSS}$ profiles obtained from the more realistic simulation
run with $N=10^5$ and starting with
$W_0=8$ (see Fig.~\ref{ratio_8_100k}), which will be deeply investigated
and discussed in a forthcoming paper.}

\section{Discussion and conclusions}
\label{concl}
In this paper we have presented the results of a number of simulations
aimed at exploring the connection between the evolution of the BSS spatial
distribution and the cluster dynamical age. In this first study we
adopted simplified initial conditions and the simulations are not meant
for a detailed and direct comparison with observational data but, 
rather, the goal was to explore the fundamental dynamical aspects driving
the evolution of the BSS spatial distribution.
Our results show a few features in nice qualitative agreement with 
observations and suggest that the dynamical mechanisms explored in this
paper provide a promising route for the interpretation and 
understanding of the empirical dynamical clock found in our previous studies.

Our main result is that, because of the effect of DF,
the BSS radial distribution develops a central peak and a minimum
independently of the initial cluster concentration.  In particular,
the semi-analytical model (which, among all the possible dynamical
processes, takes into account DF only), shows the rapid formation of a
bimodal distribution with a dip progressively moving toward the
external regions of the cluster. However, this model fails to
reproduce the formation of a long-lived central peak, which is instead
observed in all dynamically evolved clusters \citep{fe12}.  The
results obtained from (preliminary) $N$-body simulations show the
formation of a sharp and stable central peak and the development of a
dip in the BSS radial distribution, regardless of the initial $W_0$.
In spite of the noisy behavior of the snapshots, it can be stated
that a bimodal distribution is set in many cases and the size of
the dip tends to increase as a function of time until (in the latest
evolutionary phases) the distribution becomes monotonic (in full
agreement with the observations).

It is worth recalling the main differences between the two
approaches we followed: $(i)$ in the semi-analytical approach the distribution of
field stars is \emph{static}, while in the $N$-body simulations the
field component changes self-consistently with time, following the core contraction and,
especially, the halo expansion; $(ii)$ in the $N$-body system, various
collisional phenomena originating from 2-body and 3-body interactions
with \emph{small} impact parameters are acting during the entire
evolution, while the semi-analytical model takes into account only the
DF effect (i.e.~the consequence of 2-body interactions with \emph{large} 
impact parameters). Despite the higher degree of
realism of the $N$-body approach, from the dynamical point of view the
performed simulations are far from being representative of real clusters
because of both a too small number of particles ($10^4$) and an
unrealistically high fraction of heavy species (especially BSSs) with
respect to the lighter background component.  The main effect of these
limitations is to induce a too fast global evolution of the system
(cf.~Figures \ref{rlags1_8} and \ref{rlags_W8_100k}), producing
unrealistically short dynamical time-scales for the simulated
clusters, especially in the lowest concentration ($W_0=4$) case.
Thus, it is very possible that the low-mass stars (which are the main
responsible for the DF action on the test stars) in a real GC behave
much more like the static background in the semi-analytical, DF-only
approach, than in our (small) $N$-body simulations (cf.~the
long-dashed curves in Fig.s~\ref{rlags1_8} and \ref{rlags_W8_100k}).

More realistic simulations are therefore necessary to investigate this
possibility and to properly follow the time evolution of the BSS
radial distribution. In fact, while hints of a progressive outward
movement of $r_{\rm min}$ are found in some of the simulations
presented here, no reliable constraints can be obtained about the
characteristic time-scales of this process and the precise way the
shape of the dip changes with time and the external cluster regions
become devoid of BSSs.  More realistic simulations are needed also to
clarify which are exactly the internal ``engines'' of the dynamical
clock. The preliminary results presented here clearly point out that
DF is able to set the peak and the dip in the BSS
distribution. However, we still have to determine which is the
dominant phenomenon (and in what regime) able to drift $r_{\rm min}$
toward the external cluster regions (either DF only, or also the core
expansion after the CC, or further dynamical processes).  Certainly,
the presence of primordial binaries and an external tidal field should
also be taken into account because the former would presumably favor a
smoother collisional evolution of the system (by mitigating
CC) and the latter would limit the expansion of the
low-mass stars halo.  These more realistic $N$-body simulations are in
progress and will be described in forthcoming papers.

\acknowledgments This research is part of the project Cosmic-Lab
funded by the European Research Council (see
http://www.cosmic-lab.eu).
MP is grateful for support from KASI-Yonsei DRC program of Korea
Research Council of Fundamental Science and Technology (DRC-12-2-KASI),
and from NRF of Korea to CGER.
We acknowledge the CINECA award under the ISCRA initiative for
the availability of high performance computing resources and
we wish to thank Michele Trenti for his support on simulation
setup. We are also grateful to the anonymous referee for the
valuable comments and suggestions.

\appendix
\section{Dynamical friction in a Plummer distribution function}
\label{app_plum}
The semi-analytical treatment of DF of Sect.~\ref{semian} is based on a
numeric calculation of the deceleration suffered by a star on a given
orbit in a Plummer potential.  In the following we give all the
relevant details.

The distribution function leading to the Plummer model, with total mass $M$ and
characteristic radius $r_0$, is
\begin{equation}
f(r,v)=k\left[-\psi(r)-\frac{v^2}{2}\right]^{7/2},
\label{plum}
\end{equation}
with $k$ a normalization constant and
\begin{equation}
\psi(r)=-\frac{GM}{\sqrt{r^2+r_0^2}} \label{pot}
\end{equation}
the gravitational potential.
The corresponding mass density generating this potential is
\begin{equation}
\rho(r)=\frac{\rho_0r_0^5}{(r^2+r_0^2)^{5/2}}=-\frac{\psi^5\rho_0}{\sigma^{10}},
\label{a1}
\end{equation}
with $\rho_0\equiv 3M/4\pi r_0^3$ being the central density and
$\sigma\equiv (GM/r_0)^{1/2}$ a characteristic velocity.
Thus,
the integral in Eq. (\ref{grv1}) yields
\begin{equation}
g(r,v)=-4\pi 2^{3/2}k\psi^5\int_0^{v(-2\psi)^{-1/2}}w^2(1-w^2)^{7/2}\ud w=
-4\pi 2^{3/2}k\psi^5\int_u^\infty y^8(1+y^2)^{-6} \ud y
\end{equation}
where the substitutions $w=v(-2\psi)^{-1/2}$, $y=(w^{-2}-1)^{1/2}$ have been applied and
$u\equiv (-2\psi v^{-2}-1)^{1/2}$.
The last integral gives
\begin{equation}
g(r,v)=g(\psi,u)=\alpha\psi^5 \left[ \frac{1}{2}\tan^{-1}(u)+
\left(\frac{u^9}{2}-
\frac{79}{21}u^7-\frac{64}{15}u^5-\frac{7}{3}u^3-\frac{u}{2}\right)
(1+u^2)^{-5}-\frac{\pi}{4}\right]
\label{grv}
\end{equation}
with $\alpha\equiv 7\pi\sqrt{2}k/16$.

In order to determine $k$ and then $\alpha$, from Eq. (\ref{grv1}) applied to
the escape velocity ($u=0$) we note that
\begin{equation}
\rho(r)=mg(\psi,0)=-\frac{7\sqrt{2}}{64}\pi^2 m\psi^5 k.
\end{equation}
Thus, comparison with Eq. (\ref{a1}) implies that
$k=32\sqrt{2}\rho_0(7\pi^2m\sigma^{10})^{-1}$ and $\alpha=4\rho_0(\pi m\sigma^{10})^{-1}$.


\end{document}